\documentstyle[aps,preprint]{revtex}

\renewcommand{\theequation}{\arabic{equation}}
\newcommand\beq{\begin{equation}}
\newcommand\eeq{\end{equation}}
\newcommand\bea{\begin{eqnarray}}
\newcommand\eea{\end{eqnarray}}

\begin{document}


\date {\today}

\title{$CP^{1}$ model with Hopf term and fractional spin statistics}

\author{Soon-Tae Hong\footnote{electronic address:sthong@ccs.sogang.ac.kr},
  Bum-Hoon Lee\footnote{electronic address:bhl@ccs.sogang.ac.kr}, and
  Young-Jai Park\footnote{electronic address:yjpark@ccs.sogang.ac.kr}}

\address{Department of Physics and Basic Science Research Institute,\\
Sogang University, C.P.O. Box 1142, Seoul 100-611, Korea}

\maketitle

\begin{abstract}
We reconsider the $CP^{1}$ model with the Hopf term by using the 
Batalin-Fradkin-Tyutin (BFT) scheme, which is an improved version of 
the Dirac quantization method. We also perform a semi-classical quantization
of the topological charge $Q$ sector by exploiting the collective coordinates 
to explicitly show the fractional spin statistics.
\end{abstract}

\vspace{1.0cm}
\hspace{1.2 cm} Keywords: $CP^{1}$ model, Hopf term, BFT formalism

\newpage
\section{Introduction}

Since the (2+1) dimensional O(3) nonlinear sigma model (NLSM) was first 
discussed by Belavin and Polyakov~\cite{polyakov75}, there have 
been lots of attempts to improve this soliton model associated 
with the homotopy group $\pi_{2}(S^{2})=Z$.  In particular, the 
configuration space in the O(3) NLSM is infinitely connected to yield 
the fractional spin statistics, which was first shown by Wilczek and 
Zee~\cite{wilczek83} via the additional Hopf term.  The creation and 
annihilation mechanism of a Skyrmion-anti Skyrmion pair in the vacuum 
through the channel of $2\pi$ rotation of the Skyrmion was also 
studied in the O(3) NLSM~\cite{jaro85} to discuss the Hopf topological 
invariant and linking number~\cite{wilczek83}.  Moreover the O(3) 
NLSM with the Hopf term was canonically quantized~\cite{bowick86} and 
the $CP^{1}$ model with the Hopf 
term~\cite{pan88,kovner89,semenoff92,ban94,cgm01}, which can be related with 
the O(3) NLSM via the Hopf map projection from $S^{3}$ to $S^{2}$, was also 
canonically quantized later~\cite{kovner89}.  In fact, the $CP^{1}$ model has 
better features than the O(3) NLSM, in the sense that the action of the 
$CP^{1}$ model with the Hopf invariant has a desirable manifest locality, 
since the Hopf term has a local integral representation in 
terms of the physical fields of the $CP^{1}$ model~\cite{wilczek83}.  
Furthermore, this manifest locality in time is crucial for a consistent 
canonical quantization.  However, there still exist several ambiguities 
in performing the quantization rigorously and in deriving explicit expressions of 
fractional spin.~\cite{pan88,kovner89,semenoff92,ban94,cgm01}

On the other hand, physical systems with constraints was systematically 
studied by Dirac~\cite{di}, according to whom in the second 
class constraint system one needs to use the Dirac brackets instead of the 
Poisson brackets to proceed to quantize the physical system.  However, in this 
Dirac quantization scheme, we have difficulties in finding canonically 
conjugate pairs due to field operator ordering ambiguity.  To circumvent 
such problems, Batalin, Fradkin and Tyutin (BFT)~\cite{BFT,BFT1,kpr} invented a 
scheme which converts the second class constraints into first class ones by 
introducing auxiliary fields.  Recently this BFT scheme has been applied to 
several areas of current interests such as the soliton 
models~\cite{sk2,o3}, high dense matter physics~\cite{hong00q} and D-brane 
systems~\cite{hong00db}. 

The motivation of this paper is to systematically apply the BFT scheme 
\cite{BFT,BFT1,kpr}, which is an improved version of the Dirac quantization 
method, the Batalin, Fradkin and Vilkovisky (BFV) method \cite{bfv} and the 
Becci-Rouet-Stora-Tyutin (BRST) method \cite{brst} to the $CP^{1}$ 
model with the Hopf term~\cite{pan88,kovner89,semenoff92,ban94,cgm01}.  As a 
result, we will explicitly show that the $CP^{1}$ model has a fractional 
spin statistics. In section 2 we convert the second-class constraints into 
first-class ones following the BFT method to construct first-class BFT 
physical fields and directly derive the compact expression of a first-class 
Hamiltonian in terms of these fields.  We construct in section 3 a 
BRST-invariant gauge fixed Lagrangian in the BFV scheme through the standard 
path-integral procedure.  Exploiting collective coordinates, in section 4 we 
perform a semi-classical quantization to describe the fractional statistics 
explicitly. 


\section{First-class constraints and first-class Hamiltonian}
\setcounter{equation}{0}
\renewcommand{\theequation}{\arabic{section}.\arabic{equation}}


In this section, let us apply the BFT scheme to the (2+1) dimensional $CP^{1}$ 
model with a Hopf term~\cite{pan88,kovner89,semenoff92}, which is 
a second-class constraint system, and whose Lagrangian is given as 
\beq
L=\int{\rm d}^{2}x~\left[(D_{\mu}Z_{\alpha})^{*}(D^{\mu}Z_{\alpha})
+\frac{\Theta}{4\pi^{2}}\epsilon^{\mu\nu\rho}A_{\mu}
\partial_{\nu}A_{\rho}\right]
\label{laga}
\eeq
with $D_{\mu}=\partial_{\mu}-iA_{\mu}$ and $A_{\mu}$ is defined as 
\beq
A_{\mu}=-\frac{i}{2}(Z^{*}_{\alpha}\partial_{\mu}Z_{\alpha}
-Z_{\alpha}\partial_{\mu}Z^{*}_{\alpha}).
\label{amu}
\eeq
Here $Z_{\alpha}=(Z_{1},Z_{2})$ is a multiplet of complex scalar fields with a
constraint 
\begin{equation}
\Omega_{1}=Z_{\alpha}^{*}Z_{\alpha} - 1=|Z|^{2}-1\approx 0.  
\label{c1}
\end{equation}
On the other hand, this Lagrangian (\ref{laga}) can also be rewritten as the 
following standard form~\cite{ban94},  
\bea
L&=&\int{\rm d}^{2}x~\left[\partial_{\mu}Z_{\alpha}^{*}\partial^{\mu}Z_{\alpha}
-(Z_{\alpha}^{*}\partial_{\mu}Z_{\alpha})(Z_{\beta}\partial^{\mu}
Z_{\beta}^{*})+\lambda \Omega_{1}+{\cal L}_{H}\right],\nonumber\\
{\cal L}_{H}&=&-\frac{\Theta}{8\pi^{2}}\epsilon^{\mu\nu\rho}(Z^{*}_{\alpha}\partial_{\mu}Z_{\alpha}
-\partial_{\mu}Z_{\alpha}^{*}Z_{\alpha})\partial_{\nu}Z_{\beta}^{*}\partial_{\rho}Z_{\beta}.
\label{cplag}
\eea
This Lagrangian is invariant under a local U(1) gauge symmetry transformation
\beq
Z_{\alpha}(x)\rightarrow e^{i\theta (x)}Z_{\alpha}(x),
\label{u1gauge}
\eeq
and we have explicitly included the constraint $\Omega_{1}$.  By performing 
the Legendre transformation, one can obtain the canonical Hamiltonian, 
\bea
H_{c}&=&H-\int{\rm d}^{2}x~\lambda\Omega_{1}\nonumber\\
H&=&\int{\rm d}^{2}x~\left[|\Pi_{\alpha}+\frac{\Theta}{8\pi^{2}}
\Pi_{\alpha}^{H}|^{2}+|\partial_{i}Z_{\alpha}|^{2}
-(Z_{\alpha}^{*}\partial_{i}Z_{\alpha})
(Z_{\beta}\partial_{i}Z_{\beta}^{*})\right]
 \label{hc}
\eea
where $\Pi_{\alpha}$ are the canonical momenta conjugate to the complex scalar 
fields $Z_{\alpha}$ given by 
\begin{eqnarray}
\Pi_{\alpha}&=&\dot{Z}_{\alpha}^{*}-Z_{\alpha}^{*}Z_{\beta}\dot{Z}_{\beta}^{*}
-\frac{\Theta}{8\pi^{2}}\Pi_{\alpha}^{H},
\nonumber\\
\Pi_{\alpha}^{H}&=&\epsilon^{ij}(Z_{\alpha}^{*}\partial_{i}Z_{\beta}^{*}\partial_{j}Z_{\beta}
+(Z_{\beta}^{*}\partial_{i}Z_{\beta}-\partial_{i}Z_{\beta}^{*}
Z_{\beta})\partial_{j}Z_{\alpha}^{*}),
\label{cojm}
\end{eqnarray}
and $\Pi_{\alpha}^{*}$ are the complex conjugate of $\Pi_{\alpha}$.  Even 
though there is no explicit physical contribution of the $\Omega_{1}$ term to 
$H_{c}$, we have the canonical momentum conjugate to the multiplier field 
$\lambda$ to yield the first class constraint 
$\Omega_{0}=\Pi_{\lambda}\approx 0$.  The time evolution of the constraint 
$\Omega_{0}$ with $H_{c}$ yields the constraint $\Omega_{1}$ in Eq. 
(\ref{c1}), and subsequent time evolution of the constraint $\Omega_1$ yields 
an additional secondary constraint 
\begin{equation}
\Omega_{2}=Z_{\alpha}^{*}(\Pi_{\alpha}^{*}+\frac{\Theta}{8\pi^{2}}\Pi_{\alpha}^{H*})
+Z_{\alpha}(\Pi_{\alpha}+\frac{\Theta}{8\pi^{2}}\Pi_{\alpha}^{H})\approx 0.  
\label{const22}
\end{equation}
On the other hand, we could fix the multiplier field $\lambda$ such as 
$\lambda=-\frac{1}{\Delta_{12}}\{\Omega_{2},H\}$ so that we can have closed 
constraint algebra with no more time evolution of the constraints~\cite{di}.  
Since this multiplier field $\lambda$ is non-dynamical after fixing as above, 
one can decouple the irrelevant conjugate pair $(\lambda, \Pi_{\lambda})$ 
from our system of interest.  As a result, all the second-class constraints 
$\Omega_{1}$ and $\Omega_{2}$ form the following constraint algebra 
\begin{equation}
\Delta_{kk^{\prime}}(x,y)=\{\Omega_{k}(x),\Omega_{k^{\prime}}(y)\}
=2\epsilon^{kk^{\prime}}|Z|^{2}\delta(x-y)  \label{delta}
\end{equation}
with $\epsilon^{12}=-\epsilon^{21}=1$.




Next, we consider the Poisson brackets of the fields to construct the Dirac 
brackets defined as
\begin{equation}
\{A(x),B(y)\}_{D}=\{A(x),B(y)\}-\int d^2z d^2 z^{\prime}
\{A(x),\Omega_{k}(z)\}\Delta^{k k^{\prime}}
\{\Omega_{k^{\prime}}(z^{\prime}),B(y)\}
\end{equation}
with $\Delta^{k k^{\prime}}$ being the inverse of $\Delta_{k k^{\prime}}$ in
Eq. (\ref{delta}).  
After some algebraic manipulation, we obtain the commutators as follows 
\begin{eqnarray}
\{Z_{\alpha}(x),Z_{\beta}(y)\}_{D}&=&
\{Z_{\alpha}^{*}(x),Z_{\beta}(y)\}_{D}=0,  \nonumber \\
\{Z_{\alpha}(x),\Pi_{\beta}(y)\}_{D}&=&(\delta_{\alpha\beta}
-\frac{Z_{\alpha}Z_{\beta}^{*}}{2|Z|^{2}})\delta(x-y),  \nonumber \\
\{Z_{\alpha}(x),\Pi_{\beta}^{*}(y)\}_{D}&=&
-\frac{Z_{\alpha}Z_{\beta}}{2|Z|^{2}}\delta(x-y),  \nonumber \\
\{\Pi_{\alpha}(x),\Pi_{\beta}(y)\}_{D}&=&\frac{1}
{2|Z|^{2}}(\Pi_{\alpha}Z_{\beta}^{*}-Z_{\alpha}^{*}
\Pi_{\beta}-\frac{\Theta}{4\pi^{2}}(\Pi_{\alpha}^{H}Z_{\beta}^{*}-Z_{\alpha}^{*}
\Pi_{\beta}^{H}))\delta (x-y),\nonumber\\  
\{\Pi_{\alpha}(x),\Pi_{\beta}^{*}(y)\}_{D}&=&\frac{1}
{2|Z|^{2}}(\Pi_{\alpha}Z_{\beta}-Z_{\alpha}^{*}\Pi_{\beta}^{*}
-\frac{\Theta}{4\pi^{2}}(\Pi_{\alpha}^{H}Z_{\beta}-Z_{\alpha}^{*}
\Pi_{\beta}^{H*}))\delta (x-y).
\nonumber\\ 
\label{commst}
\end{eqnarray}
Note that we have the Hopf term contributions in the last two Dirac commutators in Eq. (\ref{commst}). 

Now, we calculate the symmetric energy-momentum tensor  
\begin{eqnarray}
T^{\mu\nu}&=&\partial^{\mu}Z^{*}\partial^{\nu}Z
-(Z\partial^{\mu}Z^{*})(Z^{*}\partial^{\nu}Z)
-\frac{\Theta}{8\pi^{2}}\epsilon^{\mu\rho\sigma}\left((Z^{*}\partial^{\nu}Z)
(\partial_{\rho}Z^{*}\partial_{\sigma}Z)
\right.\nonumber\\
& &\left.+(Z^{*}\partial_{\rho}Z-\partial_{\rho}Z^{*}Z)(\partial_{\sigma}Z^{*}\partial^{\nu}Z)\right)
+{\rm c.c}
\nonumber\\
& &-g^{\mu\nu}(\partial_{\sigma}Z^{*}\partial^{\sigma}Z)
   +g^{\mu\nu}(Z^{*}\partial_{\sigma}Z)(Z\partial^{\sigma}Z^{*})
\nonumber\\
& &+\frac{\Theta}{8\pi^{2}}g^{\mu\nu}\epsilon^{\alpha\beta\rho}
(Z^{*}\partial_{\alpha}Z-\partial_{\alpha}Z^{*}Z)
(\partial_{\beta}Z^{*}\partial_{\rho}Z),
\label{tt}
\end{eqnarray}
from which we can obtain the momentum operator $P^{i}$ as
\beq
P^{i}=\int{\rm d}^{2}x~T^{0i}=\int{\rm d}^{2}x~
(\Pi_{\alpha}\partial^{i}Z_{\alpha}+\Pi^{*}_{\alpha}\partial^{i}Z^{*}_{\alpha}).
\label{pi}
\eeq
This momentum operator $P^{i}$ generates the desired translation as follows 
\beq
\{P^{i},Z_{\alpha}(x)\}_{D}=\partial^{i}Z_{\alpha}(x).
\eeq
On the other hand, since the angular momentum operator $J$ is given by  
\beq
J=\int {\rm d}^{2}x \epsilon_{ij}x^{i}T^{oj},  
\label{jj}
\eeq
the rotational property of the $Z_{\alpha}$ field is obtained by treating 
the Dirac commutator 
\beq
\{J,Z_{\alpha}(x)\}_{D}=\epsilon^{ij}x^{i}\partial_{j}Z_{\alpha}(x),
\eeq
which shows that there is no anomaly term, contrast to the result of 
Ref.~\cite{kovner89}.

Following the BFT formalism \cite{BFT,BFT1,kpr}, which systematically
converts the second-class constraints into first-class ones, let us introduce
two real auxiliary fields $\Phi^{i}$ with the Poisson brackets 
$$
\{\Phi^{i}(x), \Phi^{j}(y)\}=\epsilon^{ij}\delta(x-y),
$$ 
to obtain the first-class constraints as follows
\begin{eqnarray}
\tilde{\Omega}_{1}&=&\Omega_{1}+2\Phi^{1},  \nonumber \\
\tilde{\Omega}_{2}&=&\Omega_{2}-|Z|^{2}\Phi^{2},
\label{1stconst}
\end{eqnarray}
which yield a strongly involutive first-class constraint algebra 
$\{\tilde{\Omega}_{i}(x),\tilde{\Omega}_{j}(y)\}=0$.  Here one notes that the physical fields 
$Z_{\alpha}$ are geometrically constrained to reside on the $S^{3}$ hypersphere with the modified 
norm $|Z|^{2}=1-2\Phi^{1}$.

Now, we construct the first class BFT physical fields 
$\tilde{{\cal F}}=(\tilde{Z}_{\alpha},\tilde{\Pi}_{\alpha})$ corresponding to 
the original fields ${\cal F}=(Z_{\alpha},\Pi_{\alpha})$.  These fields 
$\tilde{{\cal F}}$'s are obtained as a power 
series in the auxiliary fields $\Phi^{i}$ by demanding that they are strongly 
involutive: $\{\tilde{\Omega}_{i}, \tilde{{\cal F}}\}=0$.  After some algebra, 
we obtain the compact forms of first class physical fields as 
\begin{eqnarray}
\tilde{Z}_{\alpha}&=&Z_{\alpha}
\left(\frac{|Z|^{2}+2\Phi^{1}}{|Z|^{2}}\right)^{1/2}, 
\nonumber \\
\tilde{\Pi}_{\alpha}&=&\left(\Pi_{\alpha}-\frac{1}{2}Z_{\alpha}^{*}\Phi^{2}
\right)
\left(\frac{|Z|^{2}}{|Z|^{2}+2\Phi^{1}}\right)^{1/2}.  
\label{pitilde}
\end{eqnarray}

As discussed in Ref.~\cite{kpr}, any functional ${\cal K}(\tilde{{\cal F}})$ 
of the first class fields $\tilde{{\cal F}}$ is also first class, namely, 
$\tilde{{\cal K}}({\cal F};\Phi )={\cal K}(\tilde{{\cal F}})$.  Using this 
useful property, we easily construct a first-class Hamiltonian in terms of the 
above BFT physical variables omitting infinite iteration procedure to arrive 
at\footnote{
>From now on, for simplicity we will ignore the term proportional to 
$\Omega_{1}$, which does not yield any particular physical results.}  
\begin{equation}
\tilde{H}=\int {\rm d}x~\left[|\tilde{\Pi}_{\alpha}
+\frac{\Theta}{8\pi^{2}}\tilde{\Pi}_{\alpha}^{H}|^{2}
+|\partial_{i}\tilde{Z}_{\alpha}|^{2}
-(\tilde{Z}_{\alpha}^{*}\partial_{i}\tilde{Z}_{\alpha})(\tilde{Z}_{\beta}
\partial_{i}\tilde{Z}_{\beta}^{*})\right].
\label{htilde}
\end{equation}
We then directly rewrite this Hamiltonian in terms of the original as well as 
auxiliary fields~\footnote{In deriving the first class Hamiltonian $\tilde{H}$ 
of Eq. (\ref{hct}), we
have used the conformal map condition, $Z_{\alpha}^{*}\partial_{i}Z_{\alpha}
+Z_{\alpha}\partial_{i}Z_{\alpha}^{*}=0$.} to obtain
\bea
\tilde{H}&=&\int {\rm d}^{2}x \left[
|\Pi_{\alpha}-\frac{1}{2}Z_{\alpha}^{*}\Phi^{2}+\frac{\Theta}{8\pi^{2}R^{2}}
\Pi_{\alpha}^{H}|^{2}R\right.
\nonumber\\
& &\left.+|\partial_{i}Z_{\alpha}|^{2}\frac{1}{R}
-(Z_{\alpha}^{*}\partial_{i}Z_{\alpha})(Z_{\beta}\partial_{i}Z_{\beta}^{*})\frac{1}{R^{2}}
\right],
\label{hct}
\eea
where $R=|Z|^{2}/(|Z|^{2}+2\Phi^{1})$.  Here $\tilde{H}$ is strongly 
involutive with the first class constraints 
$\{\tilde{\Omega}_{i},\tilde{H}\}=0$.  A problem with $\tilde{H}$ in 
(\ref{hct}) is that it does not naturally 
generate the first-class Gauss law constraint from the time evolution of the 
constraint $\tilde{\Omega}_{1}$.  Therefore, by introducing an additional term 
proportional to the first class constraints 
$\tilde{\Omega}_{2}$ into $\tilde{H}$, we obtain an equivalent first class 
Hamiltonian 
\begin{equation}
\tilde{H}^{\prime}=\tilde{H}+\frac{1}{2}\int {\rm d}^{2}x \Phi^{2}
\tilde{\Omega}_{2},  
\label{hctp}
\end{equation}
which naturally generates the form invariant Gauss law constraint 
\beq
\{\tilde{\Omega}_{1},\tilde{H}^{\prime}\}=\tilde{\Omega}_{2},~\{\tilde{\Omega}_{2},\tilde{H}^{\prime}\}=0.
\eeq
Note that $\tilde{H}$ and $\tilde{H}^{\prime}$ act in the same way 
on physical states, which are annihilated by the first-class constraints. 


\section{BRST symmetries}
\setcounter{equation}{0}
\renewcommand{\theequation}{\arabic{section}.\arabic{equation}}


In this section we introduce two canonical sets of ghosts and anti-ghosts 
together with auxiliary fields in the framework of the BFV formalism 
\cite{bfv}, which is applicable to theories with the first-class 
constraints: 
\[
({\cal C}^{i},\bar{{\cal P}}_{i}),~~({\cal P}^{i}, \bar{{\cal C}}_{i}),
~~(N^{i},B_{i}),~~~~(i=1,2) 
\]
which satisfy the super-Poisson algebra 
\footnote{
Here the super-Poisson bracket is defined as 
$
\{A,B\}=\frac{\delta A}{\delta q}|_{r}\frac{\delta B}{\delta p}|_{l}
-(-1)^{\eta_{A}\eta_{B}}\frac{\delta B}{\delta q}|_{r}\frac{\delta A} {%
\delta p}|_{l} 
$ where $\eta_{A}$ denotes the number of fermions, called the ghost number, 
in $A$ and the subscript $r$ and $l$ denote right and left derivatives, 
respectively.}
\[
\{{\cal C}^{i}(x),\bar{{\cal P}}_{j}(y)\}=\{{\cal P}^{i}(x), \bar{{\cal C}}%
_{j}(y)\}=\{N^{i}(x),B_{j}(y)\}=\delta_{j}^{i}\delta(x-y). 
\]
In the $CP^{1}$ model, the nilpotent BRST charge $Q_{B}$ and the BRST invariant minimal
Hamiltonian $H_{m}$ are given by 
\begin{eqnarray}
Q_{B}&=&\int {\rm d}^{2}x~({\cal C}^{i}\tilde{\Omega}_{i}+{\cal P}^{i}B_{i}), 
\nonumber \\
H_{m}&=&\tilde{H}^{\prime}-\int {\rm d}^{2}x~{\cal C}^{1}\bar{{\cal P}}%
_{2},
\end{eqnarray}
which satisfy the relations 
\begin{equation}
\{Q_{B},H_{m}\}=0,~~Q_{B}^{2}=\{Q_{B},Q_{B}\}=0.
\end{equation}
Our next task is to fix the gauge, which is crucial to identify the BFT 
auxiliary field $\Phi^{1}$ with the St\"{u}eckelberg field.  The desired 
identification follows if one chooses the fermionic gauge fixing function 
$\Psi$ as
\begin{equation}
\Psi=\int {\rm d}^{2}x~(\bar{{\cal C}}_{i}\chi^{i}+\bar{{\cal P}}%
_{i}N^{i}),  
\end{equation}
with the unitary gauge 
\begin{equation}
\chi^{1}=\Omega_{1},~~~\chi^{2}=\Omega_{2}.
\end{equation}
Here note that the $\Psi$ satisfies the following identity
\begin{equation}
\{\{\Psi,Q_{B}\},Q_{B}\}=0.
\end{equation}

The effective quantum Lagrangian is then described as 
\begin{equation}
L_{eff}=\int {\rm d}^{2}x~(\Pi_{\alpha}^{*}\dot{Z}_{\alpha}^{*}
+\Pi_{\alpha}\dot{Z}_{\alpha}
+\pi_{\theta}\dot{\theta} +B_{2}%
\dot{N}^{2}+\bar{{\cal P}}_{i}\dot{{\cal C}}^{i}+\bar{{\cal C}}_{2} \dot{%
{\cal P}}^{2})-H_{tot}
\end{equation}
with $H_{tot}=H_{m}-\{Q_{B},\Psi\}$.  We have identified here the auxiliary 
fields $\Phi^{i}$ with a canonical conjugate pair $(\theta,\pi_{\theta})$, 
namely, 
\begin{equation}
\Phi^{i}=(\theta,\pi_{\theta}),
\end{equation}
and the terms $\int {\rm d}^{2}x~(B_{1}\dot{N}^{1}+\bar{{\cal C}}_{1}
\dot{{\cal P}}^{1})=\{Q_{B},\int{\rm d}^{2}x~\bar{{\cal C}}_{1} \dot{N}^{1}\}$ 
have been suppressed by replacing $\chi^{1}$ with $\chi^{1} +\dot{N}^{1}$.

Now let us perform path integration over the fields $B_{1}$, $N^{1}$, $\bar{%
{\cal C}}_{1}$, ${\cal P}^{1}$, $\bar{{\cal P}}_{1}$ and ${\cal C}^{1}$, by
using the equations of motion.  This leads to the effective Lagrangian of the 
form 
\begin{eqnarray}
L_{eff}&=&\int{\rm d}^{2}x~\left[\Pi_{\alpha}^{*}\dot{Z}_{\alpha}^{*}
+\Pi_{\alpha}\dot{Z}_{\alpha}+\pi_{\theta}\dot{\theta}
+B\dot{N}+\bar{{\cal P}}\dot{{\cal C}}+\bar{{\cal C}}\dot{{\cal P}}\right. 
\nonumber \\
& &\left.-|\Pi_{\alpha}-\frac{1}{2}Z_{\alpha}^{*}\Phi^{2}
+\frac{\Theta}{8\pi^{2}R^{2}}\Pi_{\alpha}^{H}|^{2}R 
+(Z_{\alpha}^{*}\partial_{i}Z_{\alpha})(Z_{\beta}\partial_{i}
Z_{\beta}^{*})\frac{1}{R^{2}}\right.
\nonumber\\
& &\left.-|\partial_{i}Z_{\alpha}|^{2}\frac{1}{R}
-\frac{1}{2}\pi_{\theta}\tilde{\Omega}_{2}+2|Z|^{2}
\pi_{\theta}\bar{{\cal C}}{\cal C}+\tilde{\Omega}_{2}N+B\Omega_{2}+\bar{{\cal P}}{\cal P}\right]
\end{eqnarray}
with the redefinitions: $N\equiv N^{2}$, $B\equiv B_{2}$, $\bar{{\cal C}}\equiv 
\bar{{\cal C}}_{2}$, ${\cal C}\equiv {\cal C}^{2}$, $\bar{{\cal P}}\equiv 
\bar{{\cal P}}_{2}$, ${\cal P}\equiv {\cal P}_{2}$.

After performing the routine variation procedure and identifying 
$N=-B+\dot{\theta}/(1-2\theta)$ we arrive at the effective 
Lagrangian of the covariant form 
\begin{eqnarray}
L_{eff}&=&\int{\rm d}^{2}x~\left[\frac{1}{1-2\theta}
(\partial_{\mu}Z_{\alpha}^{*})(\partial^{\mu}Z_{\alpha})
-\frac{1}{(1-2\theta)^{2}}
(Z_{\alpha}^{*}\partial_{\mu}Z_{\alpha})(Z_{\beta}\partial^{\mu}Z_{\beta}^{*})
\right.\nonumber\\
& &\left.+\frac{1}{(1-2\theta)^{2}}{\cal L}_{H}
-(1-2\theta)^{2}(B+2\bar{{\cal C}}{\cal C})^{2}
-\frac{1}{1-2\theta}\partial_{\mu}\theta\partial^{\mu}B
+\partial_{\mu}\bar{{\cal C}}
\partial^{\mu}{\cal C}\right]
\nonumber\\
\end{eqnarray}
which is invariant under the BRST-transformation 
\begin{eqnarray}
\delta_{B}Z_{\alpha}&=&\lambda Z_{\alpha}{\cal C},~~~ 
\delta_{B}\theta=-\lambda (1-2\theta){\cal C},  \nonumber \\
\delta_{B}\bar{{\cal C}}&=&-\lambda B,~~~ \delta_{B}{\cal C}=\delta_{B}B=0.
\end{eqnarray}
Note that at the level of Lagrangian, if we take the limit of 
${\theta}\rightarrow 0$, and integrate out decoupled auxiliary fields and 
(anti)ghost fields, one could directly recover the original unitary 
gauge-fixed Lagrangian (\ref{cplag}).


\section{Collective coordinate quantization}
\setcounter{equation}{0}
\renewcommand{\theequation}{\arabic{section}.\arabic{equation}}


In this section, we perform a semi-classical quantization of the topological 
charge $Q$ sector of the $CP^{1}$ model by exploiting the collective 
coordinates to consider physical aspects of the theory.  
 
In order to consider the quantum ground state, let us explicitly treat zero 
modes responsible for classical degeneracy by introducing desired collective 
coordinates~\cite{kovner89} satisfying the constraint $|Z|^{2}=1$ as follows 
\begin{eqnarray}
Z_{1}&=&e^{i\alpha}\cos \frac{F(r)}{2},  \nonumber \\
Z_{2}&=&e^{i\phi}\sin \frac{F(r)}{2}, 
\label{conf}
\end{eqnarray}
where $(r,\phi)$ are the polar coordinates and $\alpha (t)$ is the
collective coordinates. Here, we have used the profile function $F(r)$, 
which satisfies the boundary conditions: 
$\lim_{r\rightarrow \infty}F(r)=\pi$ and $F(0)=0$.  

It seems appropriate to comment on the collective coordinate ansatz 
(\ref{conf}), which yields 
explicit contributions of the Hopf term to the physical quantities as below.  
In general, for the case of the $CP^{N}$ model, the U(N) symmetry can be 
realized with the collective coordinate ansatz 
$Z_{\alpha}=U_{\alpha\beta}Z_{\beta}^{0}$ ($\alpha=1,\dots,N$) where 
$U_{\alpha\beta}$ are the $N\times N$ unitary 
matrices and $Z_{\alpha}^{0}$ are the hedgehog solutions such as 
$Z_{\alpha}^{0}=0$ ($\alpha<N$) and $Z_{N}^{0}=\cos \frac{F(r)}{2}$.  The 
remnant field $Z_{N+1}$  cannot be rotated to satisfy the boundary conditions on the gauge 
invariant physical quantities~\cite{kovner89} as in $Z_{2}$ in Eq. (\ref{conf}) 
for the $CP^{1}$ case.   

Using the above soliton configuration, we obtain the unconstrained
Lagrangian of the form 
\begin{equation}
L=-E+\frac{1}{2}{\cal I}\dot{\alpha}^{2}+\frac{\Theta}{2\pi}\dot{\alpha},  \label{originl}
\end{equation}
where the soliton static mass and the moment of inertia are given by 
\begin{eqnarray}
E &=&\frac{\pi}{2}\int_{0}^{\infty }{\rm d}r~r\left[ (\frac{{\rm d}F}{{\rm d}%
r})^{2}+\frac{\sin ^{2}F}{r^{2}}\right] ,  \nonumber \\
{\cal I} &=&\pi\int_{0}^{\infty }{\rm d}r~r\sin ^{2}F.
\label{inertias}
\end{eqnarray}
Introducing the canonical momentum conjugate to the proper collective 
coordinate $\alpha$ 
\begin{equation}
p_{\alpha }={\cal I}\dot{\alpha}+\frac{\Theta}{2\pi},
\end{equation}
we then have the canonical Hamiltonian as follows 
\begin{equation}
H=E+\frac{1}{2{\cal I}}(p_{\alpha }-\frac{\Theta}{2\pi})^{2}.  \label{hcc}
\end{equation}
Here one notes that, at the canonical level, there is an explicit contribution 
of the Hopf term to this Hamiltonian, which is attainable via the collective 
coordinate ansatz (\ref{conf}).

Then, substituting the configuration (\ref{conf}) into Eq. (\ref{jj}), we 
obtain the angular momentum operator of the form 
\begin{equation}
J={\cal I}\dot{\alpha}=p_{\alpha}-\frac{\Theta}{2\pi}=-i\frac{\partial}{\partial \alpha}-\frac{\Theta}{2\pi}
=-I-\frac{\Theta}{2\pi},
\label{ji}
\end{equation}
where $I$ is the isospin quantum number.  Here note that the angular momentum 
$J$ has the fractional quantum number (${\rm integer}+\frac{\Theta}{2\pi}$) 
where one can see the explicit Hopf term contribution to the total spin.  
Then, one can obtain the eigenvalues of the Hamiltonian (\ref{hcc}) as  
\begin{equation}
\langle H\rangle=E+\frac{1}{2{\cal I}}(I+\frac{\Theta}{2\pi})^{2}.  \label{jham}
\end{equation}
In fact, the above Hamiltonian can be interpreted as mass spectrum 
of a rigid rotator in the $CP^{1}$ model with the Hopf term and, for the case 
of $\Theta=\pi$, the rotator becomes a fermion, as in the (3+1) dimensional 
Skyrme soliton.  Moreover, the zero modes in the extended phase space can be 
shown to be the same as those in the original phase space with the same energy 
spectrum (\ref{jham}) as in the O(3) NLSM case~\cite{o3}.

Finally let us define the proper topological charge $Q$ associated with the 
global U(1) symmetry in terms of the canonical momenta $\Pi_{\alpha}$ as 
follows
\beq
Q=2i\int {\rm d}^{2}x~(Z_{\alpha}\Pi_{\alpha}-Z^{*}_{\alpha}\Pi^{*}_{\alpha})
=-\frac{i\Theta}{2\pi^{2}}\int {\rm d}^{2}x~\epsilon^{ij}\partial_{i}
Z^{*}_{\alpha}\partial_{j}Z_{\alpha},
\label{qtheta}
\eeq
which originates only from the Hopf term.  This Hopf term plays a crucial 
role in fermionization of the $CP^{1}$ model as the Wess-Zumino-Witten term 
does in the Skyrmion model~\cite{wzwsk}.  Note that, for the case of $\Theta=0$, one can have 
bosonic description as expected.  On the other hand, in the 
case of $\Theta=\pi$, the topological charge $Q$ can be rewritten as 
\beq
Q=\frac{1}{4\pi}\int {\rm d}^{2}x~\epsilon^{ij}F_{ij},
\label{qaf}
\eeq
where $F_{\mu\nu}=\partial_{\mu}A_{\nu}-\partial_{\nu}A_{\mu}$ with the 
definition of the gauge field $A_{\mu}$ in Eq. (\ref{amu}).  Then, the above 
expression of $Q$ is exactly the second Chern class associated with a line 
bundle with U(1)-valued transition function~\cite{yang76,semenoff92} 
and also can be expressed in terms of the topological current as follows
\beq
Q=\int {\rm d}^{2}x~B^{0},
\eeq
where the topological current is given as  
$B^{\mu}=\frac{1}{4\pi}\epsilon^{\mu\nu\rho}F_{\nu\rho}$.  By substituting 
the collective coordinates (\ref{conf}) for the unit soliton sector into Eq. 
(\ref{qtheta}), one can easily obtain 
\beq
Q=\frac{1}{2}\int_{0}^{\pi}{\rm d}F~\sin F,
\eeq
which yields unit topological charge and thus explicitly describes the fermion 
statistics.  Moreover, for the general case of $0<\Theta<\pi$ in Eq. 
(\ref{qtheta}), one can show that the topological charge operator $Q$ has 
arbitrary fractional spin statistics, which was also seen in the angular 
momentum operator (\ref{ji}).  

\section{Conclusion}

In summary, we have constructed the first-class BFT physical fields, in terms
of which the first-class Hamiltonian is formulated to be consistent with the 
Hamiltonian with the original fields and auxiliary fields.  The translational 
and rotational properties of the physical fields $Z_{\alpha}$ have been 
realized via the Dirac brackets (not the Poisson brackets) of the physical 
fields.  Since we have obtained the first-class Hamiltonian, we have 
introduced the (anti)ghost fields to obtain the BRST invariant gauge fixed 
Lagrangian and its BRST transformation rules.  On the other hand, introducing 
the collective coordinates in the soliton configuration, we have performed 
semiclassical quantization to yield the energy spectrum which can be 
interpreted as that of the rigid rotator and can also yield fractional spin 
statistics via the Hopf term contributions.  It will be interesting to 
study the $CP^{N}$ model on the noncommutative geometry through further 
investigation. 

\vspace{1cm}

\acknowledgements{
We acknowledge financial support in part from the Korea 
Research Foundation, Grant No. KRF-2001-DP0083.}

\end{document}